\documentclass[dvips]{arxstspdf}
\usepackage{flushend}
\usepackage{stfloats}
\usepackage{graphicx}

\volume{27}
\issue{1}
\pubyear{2012}
\firstpage{135}
\lastpage{143}
\doi{10.1214/09-STS278}

\makeatletter
\newcommand{\eqref}[1]{(\ref{#1})}
\newtheorem{lemma}{Lemma}
\newtheorem{proposition}{Proposition}

\newproclaim{definition}{Definition}
\makeatother

\begin{document}
\begin{frontmatter}
\vspace*{6pt}
\title{Reversing the Stein Effect}
\runtitle{Reversing the Stein Effect}

\begin{aug}
\author[a]{\fnms{Michael D.} \snm{Perlman}\corref{}\ead[label=e1]{michael@stat.washington.edu}}\and
\author[b]{\fnms{Sanjay} \snm{Chaudhuri}\ead[label=e2]{sanjay@stat.nus.edu.sg}}
\runauthor{M. D. Perlman and S. Chaudhuri}

\affiliation{University of Washington and National University of Singapore}

\address[a]{Michael D. Perlman is Professor, Department of
Statistics, University of Washington,
Seattle, Washington 98195, USA \printead{e1}.}
\address[b]{Sanjay Chaudhuri is Assistant Professor, Department of Statistics
and Applied Probability,
National University of Singapore, Singapore 117546 \printead{e2}.}

\end{aug}

%
\begin{abstract}
The \textit{Reverse Stein Effect} is identified and illustrated:
A~sta\-tistician who shrinks
his/her data toward a point chosen without reliable knowledge about the
underlying value of
the parameter to be estimated but based instead upon the observed data
will not be
protected by the minimax property of shrinkage estimators such as that
of James and Stein,
but instead will likely incur a greater error than if shrinkage were
not used.
\end{abstract}

%

\vspace*{6pt}
\end{frontmatter}

\section{The Case for Shrinkage: The Stein Effect}

Suppose that $X$ is an observed random vector in $p$-dimensional Euclidean
space $\mathbb{R}^p$ such that $X=Y+\delta$, where $\delta$ is an
unknown location parameter and $Y$ is an unobserved absolutely
continuous random
vector. Under the mild assumption that $Y\equiv X-\delta$ is
\textit{directionally symmetric},\footnote{$\vec Y\stackrel{d}{=}-\vec Y$, where
$\vec Y:=Y/\|Y\|$ is the unit
vector in the direction of $Y$ (see Appendix~\ref{appendix1}).} it is easy to
heuristically justify ``shrinkage'' estimators for $\delta$ of the form
%
\begin{equation}\label{mutilde}
\hspace*{12pt}\hat \delta_\gamma\equiv\hat \delta_\gamma(X;\delta_0)
=\gamma(X-\delta_0)\cdot(X-\delta_0)+\delta_0,
\end{equation}
where $\gamma\equiv\gamma(X-\delta_0)\in[0,1)$ and $\delta_0$ is
\textit{any} fixed shrinkage
target point in $\mathbb{R}^p$. The improvement offered by such shrinkage estimators
is often referred to as \textit{the Stein Effect.}

\begin{figure}

\includegraphics[scale=1.05]{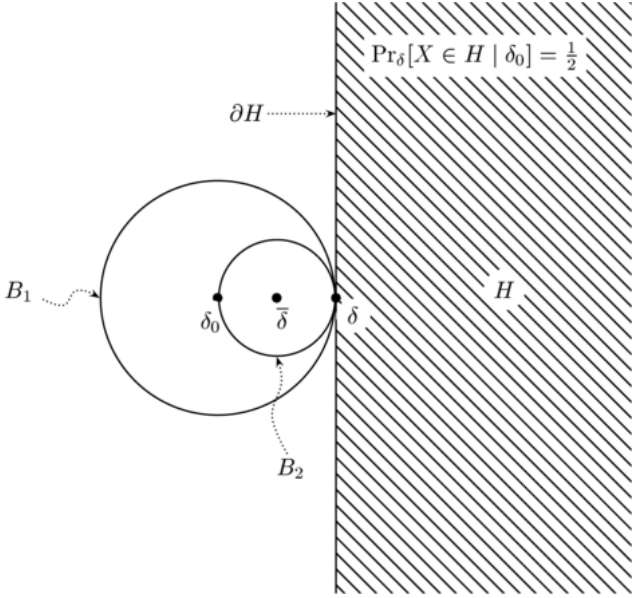}

\caption{The balls $B_1$ and $B_2$ in \protect\eqref{S1} and \protect\eqref{S2}.}\label{fig:1}
\end{figure}

First, for fixed $\delta$ and $\delta_0$, let
$B_1\equiv B_1(\|\delta-\delta_0\|;\delta_0)\subset\mathbb{R}^p$ denote the
ball of radius $\|\delta-\delta_0\|$ centered at $\delta_0$ and let
$H$ be the halfspace bounded by a hyperplane~$\partial H$ tangent to $B_1$ at
$\delta$ (see Figure~\ref{fig:1}). Then
%
\begin{eqnarray}\label{S1}
&&\{ X|\|X-\delta_0\|>\|\delta-\delta_0\|\}
=B_1^c,
\\\label{pcloser}
&&{\Pr}_\delta[ \|X-\delta_0\|>\|\delta-\delta_0\| |\delta_0]\nonumber
\\
&&\quad={\Pr}_\delta[ X\in B_1^c|\delta_0]\nonumber
\\[-8pt]\\[-8pt]
&&\quad>{\Pr}_\delta[ X\in H\mid \delta_0]\nonumber
\\
&&\quad=\tfrac{1}{2},\nonumber
\end{eqnarray}
where \eqref{pcloser} follows from directional symmetry
by Pro\-position \ref{prop1}(c) in Appendix~\ref{appendix1}. Furthermore, under
somewhat stronger but
still general assumptions (see Proposition \ref{prop2} in Appendix~\ref{appendix1}),
\begin{eqnarray}\label{S4}
&&\lim_{p\to\infty}{\Pr}_\delta[ \|X-\delta_0\|>\|\delta-\delta_0\| ] \nonumber
\\[-8pt]\\[-8pt]
&&\quad\equiv\lim_{p\to\infty}{\Pr}_\delta[X\in B_1^c]=1.\nonumber
\end{eqnarray}
Thus,
$\|X-\delta_0\|$ is usually an overestimate of
$\|\delta-\delta_0\|$, so an estimator of the form $\gamma(X-\delta_0)\cdot(X-\delta_0)$ for
$\delta-\delta_0$ should be preferable to $X-\delta_0$ itself.
Writing $\delta$ as
$(\delta-\delta_0)+\delta_0$ immediately leads to estimators for
$\delta$ of the form \eqref{mutilde}.

Second (see Appendix~\ref{appendix2}),
%
\begin{eqnarray}\label{S2}
\hspace*{15pt}\{ X\mid\exists \tilde \gamma\in[0,1)\ni\|\hat \delta_{\tilde \gamma}-\delta\|<\|X-\delta\|\}
=B_2^c,
\end{eqnarray}
where\vspace*{1pt} $\tilde \gamma\equiv\tilde \gamma(X-\delta_0,\delta-\delta_0)$
is allowed to depend on $\delta$ and $B_2\equiv B_2(\|\delta-\delta_0\|;\bar\delta)$ is the ball of radius
$\frac{1}{2}\|\delta-\delta_0\|$ centered at
$\frac{1}{2}(\delta_0+\delta)\equiv\bar\delta$. Since
$B_2^c\supset B_1^c$, also
%
\begin{equation}\label{anotherprob1}
{\Pr}_\delta[X\in B_2^c\mid\delta_0]>\tfrac{1}{2}
\end{equation}
and, under the assumptions of Proposition \ref{prop2} in Appendix~\ref{appendix1},
%
\begin{equation}\label{anotherprob2}
\lim_{p\to\infty}{\Pr}_\delta[X\in B_2^c]=1.
\end{equation}
This shows that if $\delta$ were known, then usually \textit{some}
shrinkage factor $\tilde \gamma$
applied to $X-\delta_0$ will move $X$ closer to $\delta$, again
suggesting a search for
estimators of the form \eqref{mutilde}.

\section{The Stein Paradox}\label{section2}

Assume now that $Y\sim N_p(0,\sigma^21_p)$, the multivariate normal
distribution with mean
0 and covariance matrix $\sigma^21_p$, where $\sigma^2>0$ is known,
so $X\sim N_p(\delta,\sigma^21_p)$. In this simple case, the James--Stein (JS)
estimator for $\delta$ is given by
\begin{eqnarray}\label{JSest}
\hat \delta_{\mathrm{JS}}&\equiv&\hat \delta_{\mathrm{JS}}(X;\delta_0)\nonumber
\\[-8pt]\\[-8pt]
&=&\biggl(1-\frac{\sigma^2(p-2)}{\|X-\delta_0\|^2}\biggr)(X-\delta_0)+\delta_0,\nonumber
\end{eqnarray}
where $\delta_0$ is a
fixed but \textit{arbitrary} point in $\mathbb{R}^p$. The truncated
$\equiv$ ``plus-rule'' JS
estimator
\begin{eqnarray}\label{JStrunc}
\hat \delta_{\mathrm{JS}}^+&\equiv&\hat \delta_{\mathrm{JS}}^+(X;\delta_0)\nonumber
\\[-8pt]\\[-8pt]
&=&\biggl(1-\frac{\sigma^2(p-2)}{\|X-\delta_0\|^2} \biggr)^+(X-\delta_0)+\delta_0\nonumber
\end{eqnarray}
is a shrinkage estimator of the form \eqref{mutilde}. These re\-nowned
estimators have the
property that when\break \mbox{$p\ge3$}, they dominate $X$ under both the mean
square error ($\operatorname{MSE}$) and
Pitman closeness (PC) criteria:\footnote{See Baranchik (\citeyear{Baranchik1964}) or
Efron and Morris (\citeyear{EfronMorris1973}) for \eqref{baranchik}, James and Stein (\citeyear{JamesStein1961}),
Efron and Morris (\citeyear{EfronMorris1973}),
Arnold (\citeyear{Arnold1981}),
Anderson (\citeyear{Anderson1984}),
Berger (\citeyear{Berger1985}), or
Lehmann and Casella (\citeyear{LehmannCasella1998}) for
\eqref{MSEcomparison}, our
Appendix~\ref{appendix3} for \eqref{S12}, and
Efron (\citeyear{Efron1975}) or
Sen, Kubokawa and Saleh (\citeyear{SenKubokawaSaleh1989}) for \eqref{Efroneq}.
In Efron's equation (2.11), page 265, the second inequality
should be reversed.} for every fixed $\delta,\delta_0\in\mathbb{R}^p$,
%
\begin{eqnarray}\label{baranchik}
\hspace*{22pt}&&\mathrm{E}_\delta[\|\hat \delta_{\mathrm{JS}}^+(X;\delta_0)-\delta\|^2|\delta_0]\nonumber
\\[-8pt]\\[-8pt]
&&\quad< \mathrm{E}_\delta[\|\hat \delta_{\mathrm{JS}}(X;\delta_0)-\delta\|^2|\delta_0]\nonumber
\\\label{MSEcomparison}
&&\quad< \mathrm{E}_\delta[\|X-\delta\|^2]\equiv p\sigma^2,
\\
&&{\Pr}_\delta[ \|\hat \delta_{\mathrm{JS}}^+(X;\delta_0)-\delta\|<\|X-\delta\||\delta_0]\nonumber\label{S12}
\\[-8pt]\\[-8pt]
&&\quad> {\Pr}_\delta[ \|\hat \delta_{\mathrm{JS}}(X;\delta_0)-\delta\|<\|X-\delta\||\delta_0]\nonumber
\\\label{Efroneq}
&&\quad= {\Pr}\biggl[\chi_p^2\biggl(\frac{\|\delta-\delta_0\|^2}{4\sigma^2}\biggr)
\ge\frac{\|\delta-\delta_0\|^2}{4\sigma^2}+\frac{p-2}{2}\biggr]
\\\label{Efronineq}
&&\quad>\frac{1}{2}
\end{eqnarray}
and approaches 1 as $p\to\infty$ if $\frac{\|\delta-\delta_0\|}{\sigma}=o(p)$ (apply
Chebyshev's inequality), where $\chi_p^2(\eta)$ denotes a noncentral
chi-square random variate with $p$ degrees of freedom and noncentrality parameter
$\eta$. Note especially that:
\begin{enumerate}[(A)]
\item[(A)] the improvements offered by the JS estimators can be great,
especially when
$p$ is large: if \mbox{$\delta=\delta_0$}, then
$\operatorname{MSE}(\hat \delta^+_{\mathrm{JS}})<\operatorname{MSE}(\hat \delta_{\mathrm{JS}})
=2\sigma^2\ll p\sigma^2$, and
if
$\|\delta-\delta_0\|=o(p)$ with $\sigma^2$ fixed, then
${\Pr}_\delta[ \|\hat \delta-\delta\|<\|X-\delta\|]\to1$ as
$p\to\infty$ for both $\hat \delta=\hat \delta_{\mathrm{JS}}$ and $\hat
\delta_{\mathrm{JS}}^+$;
\item[(B)] the $\operatorname{MSE}$ and PC dominances of $X$ by $\hat \delta_{\mathrm{JS}}$ and
$\hat \delta_{\mathrm{JS}}^+$ hold even if the true mean $\delta$ is arbitrarily
far from the shrinkage
target $\delta_0$.
\end{enumerate}

Of the two properties (A) and (B), it is (B) that is most surprising,
since it is not difficult to construct estimators that satisfy (A), for example,
a~Bayes estimator w.r. to a normal prior centered at~$\delta_0$. However, such
a Bayes estimator will not satisfy~(B), the difference stemming from the fact
that the Bayes estimator will have a constant shrinkage factor, while the shrinkage
factors in \eqref{JSest} and \eqref{JStrunc} are adaptive.\footnote{In fact,
the JS estimator can be derived via an empirical Bayes argument based on such priors---see
Stein [(\citeyear{Stein1966}), page 356],
Efron and Morris [(\citeyear{EfronMorris1973}), pages 117--118],
Arnold [(\citeyear{Arnold1981}), Section~11.4].}

When first discovered, the domination of $X$ by the JS estimators was
highly surprising, because the estimator~$X$ itself is as follows:\footnote{Cf. Berger (\citeyear{Berger1985}),
Lehmann and Casella (\citeyear{LehmannCasella1998}).}
\begin{longlist}[(a)]
\item[(a)] the best unbiased estimator of $\delta$,
\item[(b)] the best translation-invariant estimator of $\delta$,
\item[(c)] the maximum likelihood estimator (MLE) of~$\delta$,
\item[(d)] a minimax estimator of $\delta$, and
\item[(e)] an admissible estimator of $\delta$ when $p=1$ or 2.
\end{longlist}
So compelling were these properties of $X$ that its domination by the
JS estimators came to be
known as \textit{the Stein Paradox}.\footnote{Cf. Efron and Morris (\citeyear{EfronMorris1977}).}

\section{Lost in Space: The Reverse Stein Effect}
\textit{Star Trek, Stardate 4598.0:} The Federation Starship U.S.S.
Enterprise, about to rendezvous with interstellar space station Delta, was struck by a
mysterious distortion of the space-time continuum that disrupted all its power systems,
including navigation, communications, and computers. Out of control, the Enterprise careened
wildly and randomly through interstellar space at maximum warp for\break three days
until, equally mysteriously, its warp drive went off-line and the ship came to a full
stop. Captain Kirk knew that, without power and communication, their only hope for rescue
was to launch a~probe that would come close enough to Delta to be detected and convey their present
location.

By means of stellar charts, Lieutenant Ohura determined the present
location X of the Enterprise, but because all computer records had been lost, the
location~$\delta$ of station Delta was unknown. Mr. Chekov, fresh out of the Space Academy
where he studied multivariate statistical analysis under Admiral Emeritus Stein,
immediately suggested a solution:

``We can utilize the Stein Effect! Because the Enterprise essentially
followed a random walk while out of control we know that
$X\sim N_3(\delta,\sigma^21_3)$,
while from the duration of the disruption and the characteristics of
our warp engines we know
that $\sigma=2400$ light-years. If we use the truncated James--Stein estimator
$\hat \delta_{\mathrm{JS}}^+(X;\delta_0)$ with $p=3$ to estimate $\delta$ by
shrinking~$X$ toward a
fixed point $\delta_0$, then by \eqref{MSEcomparison} and \eqref{Efronineq},
$\hat \delta_{\mathrm{JS}}^+(X;\delta_0)$ is more likely to be closer to Delta
than our present location~$X$ is, no matter where Delta is! And what's more, we can shrink $X$
toward any $\delta_0$ that we like!''

``Amazing!'' Kirk said. ``Now I wish I had paid more attention in my stats
class,'' (\textit{smiling to himself: but that's not how one makes
Admiral!}) ``But what about
$\delta_0$? To what shrinkage target point\break should we actually send our
probe?''

``Why, toward Earth, of course,'' Scotty\footnote{A.k.a. James
Doohan, who, during the
writing of this paper, beamed out of this universe on July 20, 2005,
the 36th anniversary of
the first human landing on an extraterrestrial body.} said in his thick
Scottish brogue.
``The Scotch there is the best in the galaxy.''

``No, toward Qo'noS\footnote{The Capitol of the
Klingon Empire.}'' Lt. Worf\footnote{Yes, we know, Worf didn't
appear until \textit{Star
Trek: The Next Generation}---some slack, please.} exclaimed. ``Perhaps
they will send us
some fresh qagh\footnote{A Klingon dish of serpent worms, best when
served live.}---I am
\textit{so} tired of this replicated stuff.''

``Permit me to suggest Denobula,''
Dr. Phlox\footnote{Okay, he appeared a century earlier on \textit{Star Trek: Enterprise}---more slack please.}
offered. ``Tomorrow is the tenth wedding anniversary of my third wife and her fourth husband---perhaps
the probe might convey
my congratulations to them.''

Suggestions for the shrinkage target point $\delta_0$ were soon
received from every member of
the 400-person crew, all except Mr. Spock. After several minutes he
raised his left eyebrow and
said ``This is not logical. Please accompany me to the
holodeck.\footnote{And still more slack.}''

When the officers were assembled on the holodeck, Spock commanded:
``Computer,\footnote{Ok, let's suppose that the computer power has
been restored, but only
momentarily.} construct a three-dimensional star chart showing the
distribution in the
galaxy of the homeworlds $\delta_0$ of our crew members. What if any
statistical properties
does this distribution possess?''

``The dis-tri-bu-tion of home-worlds is such that $\delta_0$ is
di-rec-tion-al-ly sym-me-tric a-bout our pre-sent lo-ca-tion $X$,'' the
computer intoned monotonically.

\begin{figure}

\includegraphics{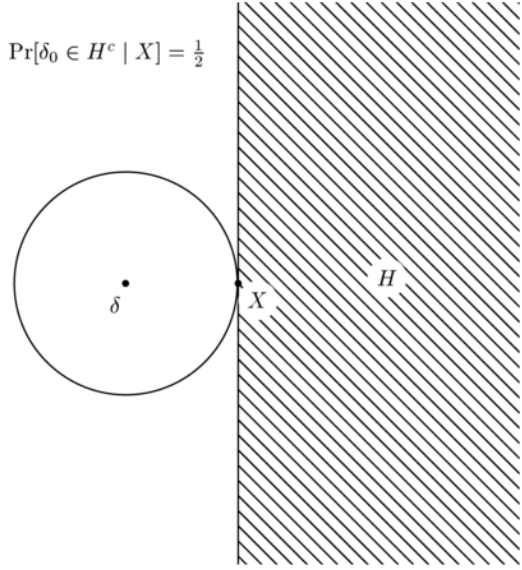}

\caption{The complement of $H$ is the set \protect\eqref{H3}.}\label{fig:2}
\end{figure}

``Computer, display the following set:
%
\begin{equation}\label{H3}
\{ \delta_0\mid\exists\check\gamma\in[0,1)\ni\|\hat \delta_{\check\gamma}-\delta
\|<\|X-\delta\|\},
\end{equation}
where $\check\gamma\equiv\check\gamma(X-\delta_0,X-\delta)$
may depend on $\delta$.''

``This set is ex-act-ly $H^c$, the com-ple-ment of the closed
half-space $H$ in Figure~\ref{fig:2} on my mon-i-tor.''

``Then, since ${\Pr}[\delta_0\in H^c|X ]=\frac{1}{2}$ by
directional\break
symmetry, this shows that shrinkage toward a randomly chosen $\delta_0$
would have at most a 50--50 chance of moving $X$ closer to $\delta$
even when the
shrinkage factor is chosen optimally for $\delta$.''

``As for James--Stein shrinkage,'' Spock continued, ``Computer, for representative
values of $\delta$, display the set of all $\delta_0$ such that the
James--Stein shrinkage
estimator $\hat \delta_{\mathrm{JS}}^+(X;\delta_0)$ lies closer to $\delta$ than
does our present location
$X$.''

\begin{figure}
\begin{tabular}{@{}c@{}}

\includegraphics{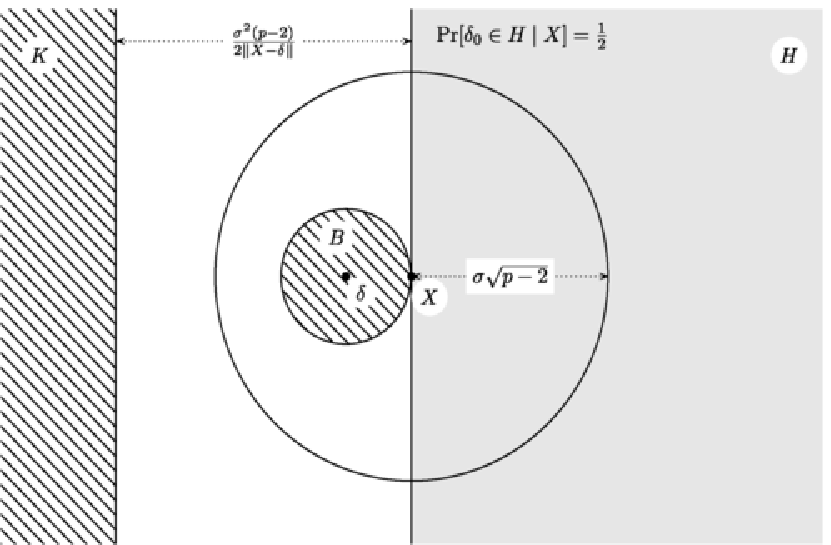}
\\
(a) $\|X-\delta\|<\frac{\sigma}{2}\sqrt{p-2}$\\[6pt]

\includegraphics{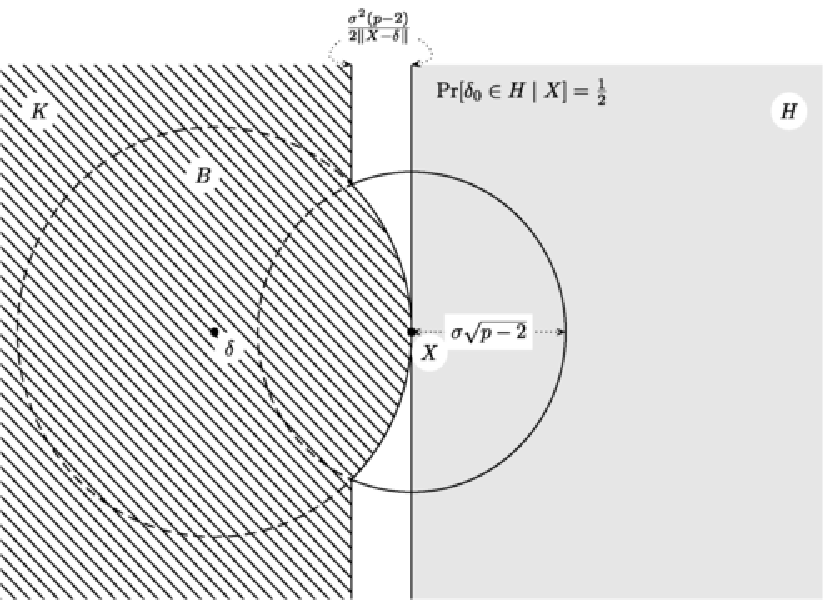}
\\
(b) $\|X-\delta\|>\frac{\sigma}{2}\sqrt{p-2}$\\
\end{tabular}
\caption{The cross-hatched region is the set $\{\delta_0|\|\hat
\delta^+_{\mathrm{JS}}(X;\delta_0)-\allowbreak\delta\|<\|X-\delta\|\}$.}\label{S3333}
\end{figure}

``The two re-pres-ent-a-tive ca-ses are now dis-played in Fig-ures 3a and
3b\footnote{See Appendix~\ref{appendix4} for their derivation.} on my mon-i-tor.''

``Thank you, Computer. It is apparent from these two displays,'' Spock
said to the assembled
officers, ``that the set\footnote{This set is the complement of
the cross-hatched region in Figure~\ref{S3333}a or~\ref{S3333}b} of $\delta_0$ such that
James--Stein shrinkage toward
$\delta_0$ does more harm than good is quite extensive. Furthermore,
since Mr. Chekov assures
us that this choice can be made arbitrarily, in the interest of
fairness, we may as well choose
$\delta_0$ at random from our crew members' homeworlds. But then,
contrary to Mr. Chekov's
assertion, $\hat \delta_{\mathrm{JS}}^+(X;\delta_0)$ is \textit{less likely} to be
closer to $\delta$ than
is our present location~$X$.''

``More precisely, by the directional symmetry of~$\delta_0$ about $X$, it follows from Figures~\ref{S3333}a and~\ref{S3333}b
that
\begin{eqnarray}\label{pcloser3}
&&{\Pr}[ \|\hat \delta^+_{\mathrm{JS}}(X;\delta_0)-\delta\|>\|X-\delta\||X ]\nonumber
\\[-8pt]\\[-8pt]
&&\quad> {\Pr}[ \delta_0\in H\mid X ]=\tfrac{1}{2}.\nonumber
\end{eqnarray}
If $\delta_0$ is actually symmetrically distributed about $X$, then it
is easy to
see that
%
\begin{equation}\label{condexp}
\mathrm{E}[ \hat \delta_{\mathrm{JS}}^+(X;\delta_0)\mid X]=X,
\end{equation}
so by Jensen's inequality,
\begin{eqnarray}\label{MSEcomparison3}
\hspace*{10pt}\mathrm{E}[ \|\hat \delta_{\mathrm{JS}}^+(X;\delta_0)-\delta\|^2| X]
&>& \mathrm{E}[ \|X-\delta\|^2| X]\nonumber
\\[-8pt]\\[-8pt]
&\equiv& p\sigma^2\quad\forall\delta\in\mathbb{R}^p.\nonumber
\end{eqnarray}
Furthermore, under additional but
still general assumptions,\footnote{See Proposition \ref{prop3} in
Appendix~\ref{appendix4}.}
%
\begin{equation}\label{S19}
\hspace*{15pt}\lim_{p\to\infty}{\Pr}_\delta[ \|\hat \delta^+_{\mathrm{JS}}(X;\delta_0)-\delta\|>\|X-\delta\| ]=1.
\end{equation}
Thus, it is likely that James--Stein shrinkage will actually move us
farther away from $\delta$. I conclude, therefore, that we should
simply tether the probe to the Enterprise and hope that Delta can detect our present location
$X$.''\vadjust{\goodbreak}

``Boy, Spock, you \textit{are} a party pooper,'' Bones\footnote{Dr.
McCoy.} said. ``I sure hope we don't
shrink toward Vulcan.''

``Resistance is futile,'' said Seven-of-Nine.\footnote{Right again, but how could we leave her out?}

``But, but,---I don't understand this,'' Chekov\break stammered.
``How can the James--Stein estimator be inferior to $X$ after all? Don't \eqref{pcloser3} and
\eqref{MSEcomparison3} contradict~\eqref{Efronineq} and \eqref{MSEcomparison}? For
example, under \textit{any} probability distribution for $\delta_0$,
\eqref{MSEcomparison}
yields
\begin{eqnarray}\label{MSEcomparisonunconditional}
\mathrm{E}_\delta[ \|\hat \delta_{\mathrm{JS}}^+(X;\delta_0)-\delta\|^2]
&<& \mathrm{E}_\delta[ \|X-\delta\|^2]\nonumber
\\[-8pt]\\[-8pt]
&\equiv& p\sigma^2\quad\forall\delta\in\mathbb{R}^p,\nonumber
\end{eqnarray}
while \eqref{MSEcomparison3} yields
\begin{eqnarray}\label{MSEcomparison6}
\mathrm{E}_\delta[ \|\hat \delta_{\mathrm{JS}}^+(X;\delta_0)-\delta\|^2]
&>& \mathrm{E}_\delta[ \|X-\delta\|^2]\nonumber
\\[-8pt]\\[-8pt]
&\equiv& p\sigma^2\quad\forall\delta\in\mathbb{R}^p.\nonumber
\end{eqnarray}
I am so confused!''

``Beam me to the bar, Scotty,'' Kirk finally mumbled. ``Maybe I can
figure this out after I
belt down a few.''

\section{To Shrink or Not to Shrink---That Is the Question}

Mr. Spock quickly assured Mr. Chekov that no formal contradiction
had occurred: the probabilities and expectations appearing in \eqref{MSEcomparison},
\eqref{Efronineq}, \eqref{pcloser3}, and~\eqref{MSEcomparison3} are
conditional probabilities and conditional expectations with different conditioning variables.
Furthermore, the joint distributions of $(X,\delta_0)$ in~\eqref{MSEcomparisonunconditional} and \eqref{MSEcomparison6} are
different, having joint
pdfs of the forms $f_\delta(X)f(\delta_0)$ and $f_\delta(X)f(\delta_0|X)$,
respectively. In the former, $X$ and $\delta_0$ are independent, whereas in the latter,
$\delta_0$ is dependent
on $X$.

However, Captain Kirk's dilemma\footnote{Captain Kirk is ``exactly
in the position of
Buridan's ass,'' as described in Barnard's discussion of the noninvariant
nature of the James--Stein estimator in Stein [(\citeyear{Stein1962}), page 288]. The ass,
when presented with two bales of hay, equidistant to his right and left, refused to move,
seeing no reason to prefer one direction over the other. Like Barnard, we maintain that, in
the absence of additional influences, such as prior information about the
delectability of dextral vs. sinistral hay (or a loss function reflecting a negative effect of
starvation), the ass's refusal to budge was correct.} remains: to shrink or not to shrink? If,
according to property~(B), the shrinkage target $\delta_0$ can be chosen arbitrarily
and still reduce the $\operatorname{MSE}$ and PC, can choosing
$\delta_0$ at random in some symmetric manner actually increase the
$\operatorname{MSE}$ and PC?

The short answer is yes, the Reverse Stein Effect is just as real as
the original Stein Effect itself---both are simply manifestations of the strong
curvature of spheres in multi-dimensional Euclidean space. Figures~\ref{S3333}a,~\ref{S3333}b, and the results
\eqref{pcloser3}, \eqref{MSEcomparison3}, and \eqref{S19} show
that, without some prior
knowledge of the location~$\delta$, \textit{Captain Kirk should not shrink
$X$.} If the \mbox{shrinkage} target $\delta_0$ is chosen without reliable
prior information but
instead is based upon the data $X$, the minimax/Bayesian robustness
property (B) of the JS
estimator is lost and no longer guarantees that shrinking is not
harmful on average.

The implications for statistical practice are apparent. A shrinkage
estimator is only as good
as, \textit{but no better than}, the prior information upon which it is
based. Without reliable
\textit{prior}, as opposed to \textit{posterior},\footnote{As
represented, for example, by
``data-dependent'' priors.} information, \textit{shrinkage is likely to
decrease the accuracy of
estimation.} As Barnard\footnote{Cf. Stein [(\citeyear{Stein1962}), page 288].}
concluded, if the statistical
estimation problem is truly invariant under translation, then the best
invariant estimator should be used, namely, $X$ itself.


\begin{appendix}
\section{\texorpdfstring{Directional and Spherical Symmetry; Verification~of~(\protect\ref{S4})}
{Directional and Spherical Symmetry; Verification of (4)}}
\label{appendix1}

\renewcommand{\theequation}{\arabic{equation}}
\setcounter{equation}{21}

\begin{definition}\label{Def1}
$Y\in\mathbb{R}^p$ is \textit{directionally symmetric} if
$\vec Y\stackrel{d}{=}-\vec Y$, where $\vec Y:=\frac{Y}{\|Y\|}$ is the unit
vector in the direction
of $Y$. $Y$ is \textit{directionally symmetric about} $y_0$ if $Y-y_0$ is
directionally symmetric.
\end{definition}

Clearly $Y$ is directionally symmetric if $Y$ is \textit{symmetric:}
$Y\stackrel{d}{=}-Y$.
Thus, any multivariate normal or elliptically contoured random vector
$Y$ centered at~0 is directionally symmetric. Directional symmetry is much weaker
than symmetry, as seen from
the following result.

\begin{proposition}\label{prop1}
Let $Y$ be an absolutely continuous random vector in $\mathbb{R}^p$.
The following are
equivalent:
\begin{longlist}[(c)]
\item[(a)] $Y$ is directionally symmetric.
\item[(b)] ${\Pr}[ Y\in C ]={\Pr}[ -Y\in C ]$ for every
closed convex cone
$C\subseteq\mathbb{R}^p$.
\item[(c)] ${\Pr}[ Y\in H ]=\frac{1}{2}$ for every \textup{central} (i.e., $0\in\partial H$) halfspace
$H\subseteq\mathbb{R}^p$.
\end{longlist}
\end{proposition}

\begin{pf}
The implications \textup{(a)} $\Leftrightarrow$ \textup{(b)} $\Rightarrow$ \textup{(c)} are straightforward.
We will show that \textup{(c)}
$\Rightarrow$ \textup{(a)}. Let~$P$ (resp., $Q$) denote the probability\vspace*{1pt}
distribution of $Y$ (resp., $\vec Y$). First note that since
$P[\partial H]=0$, $P[H]=\frac{1}{2}$ is equivalent to
%
\begin{equation}\label{app1}
P[ H]=P[ H^c]=P[ -H].
\end{equation}
Thus, for any two central halfspaces $H$ and $H_0$,
\begin{eqnarray*}
&&P[ H\cap H_0]-P[ H^c\cap H_0^c]
\\
&&\quad=P[ H ]-P[ H_0^c]
\\
&&\quad=P[ H^c]-P[ H_0]
\\
&&\quad=P[ H^c\cap H_0^c]-P[ H\cap H_0],
\end{eqnarray*}
hence,
\begin{eqnarray}\label{app3}
P[ H\cap H_0]&=&P[ H^c\cap H_0^c]\nonumber
\\[-8pt]\\[-8pt]
&=&P[ (-H)\cap(-H_0)].\nonumber
\end{eqnarray}
It follows from Lemma~\ref{lem1} below that
%
\begin{equation}\label{app4}
Q[ A\cap S_0]=Q[ (-A)\cap(-S_0)]
\end{equation}
for every Borel set $A\subseteq{\mathcal S}^p$ (the unit sphere in
$\mathbb{R}^p$), where
$S_0=H_0\cap{\mathcal S}^p$. Thus,
%
\begin{eqnarray}\label{app5}
\hspace*{25pt}Q[ A ]&=&Q[ A\cap S_0]+Q[ A\cap(-S_0)]\nonumber
\\
&=&Q[ (-A)\cap(-S_0)]+Q[ (-A)\cap(S_0)]
\\
&=&Q[ -A ]\nonumber
\end{eqnarray}
for every such $A$, hence, \textup{(a)} holds.
\end{pf}
%

\begin{lemma}\label{lem1}
Let $Y$ be an absolutely continuous random vector in
$\mathbb{R}^p$ and let $P$ and $Q$ be as defined above. Suppose that
$H_0\subset\mathbb{R}^p$
is a central halfspace such that $P[H_0]=\frac{1}{2}$, so also
$Q[S_0]=\frac{1}{2}$ where
$S_0=H_0\cap{\mathcal S}^p$. If
%
\begin{equation}\label{app6}
Q[ S\mid S_0]=Q[ -S\mid-S_0]
\end{equation}
for every hemisphere $S\subset{\mathcal S}^p$, then
%
\begin{equation}\label{app7}
Q[ A\mid S_0]=Q[ -A\mid-S_0]
\end{equation}
for every Borel set $A\,{\subseteq}\,{\mathcal S}^p$, which is equivalent to~\eqref{app4} because
$Q[\pm S_0]=P[\pm H_0]=\frac{1}{2}$. Since every hemisphere
$S$ has the form $H\cap{\mathcal S}^p$ for some central halfspace $H$,
\eqref{app6} is equivalent to
%
\begin{equation}\label{app8}
P[ H\mid H_0]=P[ -H\mid-H_0]
\end{equation}
for every central halfspace $H\subset\mathbb{R}^p$, which in turn is
equivalent to
\eqref{app3}.
\end{lemma}

\begin{pf}
Without loss of generality, set $H_0=\break \{(y_1,\dots,y_{p-1},y_p)\mid y_p>0\}$ so
%
\begin{equation}\label{appSzero}
S_0=\Biggl\{(y_1,\ldots,y_{p-1},y_p) \Big|\sum_{i=1}^p
y_i^2=1,y_p>0\Biggr\},\hspace*{-32pt}\vadjust{\goodbreak}
\end{equation}
and let $\pi$ denote the stereographic projection\footnote{Cf. Ambartzumian [(\citeyear{Ambartzumian1982}), page 26],
Watson [(\citeyear{Watson1983}), page 23].} of $S_0$ onto its tangent hyperplane
$L_0\equiv\{(y_1,\dots,y_{p-1},1)\}$. Then the relation
%
\begin{equation}\label{app9}
\pi(S\cap S_0)=K
\end{equation}
determines a bijection between the sets of all hemispheres $S\subset{\mathcal S}^p$ and \textit{all}
(not necessarily central) halfspaces $K\subset L_0$.

Let $\tilde {Q}$ denote the probability measure on ${\mathcal S}^p$ given by
%
\begin{equation}\label{app10}
\tilde {Q}[ A ]=Q[ -A\mid-S_0 ],
\end{equation}
so \eqref{app6} states that
%
\begin{equation}\label{app11}
Q[ S\mid S_0]=\tilde {Q}[ S ]
\end{equation}
for every Borel set $A\subseteq{\mathcal S}^p$. Let $R$ and $\tilde {R}$ denote
the probability
measures induced on $L_0$ by $Q[ \cdot\mid S_0]$ and~$\tilde {Q}$,
respectively, under the mapping
$\pi$, that is,
%
\begin{eqnarray}
R[ B ]&=&Q[ \pi^{-1}(B)\mid S_0],\label{app12}
\\
\tilde {R}[ B ]&=&\tilde {Q}[ \pi^{-1}(B) ]\label{app13}
\end{eqnarray}
for every Borel set $B\subseteq L_0$. Then for each halfspace $K\subset L_0$,
%
\begin{eqnarray}
R[ K ]&=&Q[ \pi^{-1}(K)\mid S_0]=Q[ S\cap S_0\mid S_0]\nonumber
\\[-8pt]\\[-8pt]
&\equiv& Q[S\mid S_0],\nonumber
\\
\tilde {R}[ K ]&=&\tilde {Q}[ \pi^{-1}(K)]\nonumber
\\[-8pt]\\[-8pt]
&=&\tilde {Q}[ S\cap S_0]\equiv Q[-(S\cap S_0)\mid-S_0]\nonumber
\\
&=&Q[ -S\mid-S_0]\equiv\tilde {Q}[ S ],
\end{eqnarray}
hence, $R[ K ]=\tilde {R}[ K ]\  \forall K$ by \eqref{app11}. Thus
by the Cra\-m\'{e}r--Wold device
[cf. Billingsley (\citeyear{Billingsley1982}), page~334], $R[B]=\tilde {R}[B]\  \forall B$, hence,
setting $A=\pi^{-1}(B)$ in
\eqref{app12} and~\eqref{app13},
$Q[A\mid S_0]=\tilde {Q}[A]\ \forall A$, which establi-\break shes~\eqref{app7}.
\end{pf}
%

\begin{definition}\label{Def2}
$Y\in\mathbb{R}^p$ is \textit{spherically symmetric}~$\equiv$ \textit{orthogonally invariant} if
$Y\stackrel{d}{=}\Gamma Y$ for every orthogonal transformation $\Gamma
$ of $\mathbb{R}^p$.
$Y$ is \textit{spherically symmetric about} $y_0$ if $Y-y_0$ is
spherically symmetric.
\end{definition}

For example, $Y\sim N_p(0,\sigma^21_p)$ is spherically symmetric.
Clearly spherical symmetry \vspace*{1pt}
implies symmetry. It is well known that $Y$ is spherically symmetric
iff $\vec Y$ is
uniformly distributed on the unit sphere $\mathcal{S}_p$ and is
independent of $\|Y\|$.
We now use this fact to verify \eqref{S4} by the following
proposition, where
$\delta$, $\delta_0$, $X$, $Y$, $\sigma$, $\psi$, and $\tau$ all
depend on $p$.

\begin{proposition}\label{prop2}
Assume that:
\begin{longlist}[(iii)]
\item [(i)]$\delta_0$ is (fixed or) random is independent of $X$;
\item [(ii)]$Y\equiv X-\delta$ is \textit{spherically symmetric};
\item [(iii)]
$\frac{\|\delta_0-\delta\|}{\|Y\|}\equiv\frac{\|\delta_0-\delta\|}{\|X-\delta\|}=o(p^{1/2})$\vspace*{1pt}
in probability as\break $p\to\infty$.
Then [cf. \eqref{S4}]
\begin{eqnarray}\label{newlimit5}
&&\lim_{p\to\infty}{\Pr}_\delta[ \|X-\delta_0\|>\|\delta-\delta_0\| ]\nonumber
\\[-8pt]\\[-8pt]
&&\quad\equiv\lim_{p\to\infty}{\Pr}_\delta[ X\in B_1^c]=1.\nonumber
\end{eqnarray}
The boundedness assumption \textup{(iii)} is satisfied, for example, if
$X\sim N_p(\delta,\sigma^21_p)$ and $\delta_0\sim N_p(\psi,\tau^21_p)$ with
$\|\psi-\delta\|/\sigma=o(p)$ and $\tau/\sigma=o(p^{1/2})$.
\end{longlist}
\end{proposition}

\begin{pf}
Let $\mu_p$ denote the uniform probability measure on $\mathcal{S}_p$. By \textup{(i)} and \textup{(ii)},
${\Pr}_\delta[ X\in B_1\mid\delta_0]$ depends on
$\delta_0$ only via $\|\delta_0-\delta\|$ (the radius of $B_1$), and
%
\begin{eqnarray}
\hspace*{15pt}&&{\Pr}_\delta[ X\in B_1|\|\delta_0-\delta\| ]
\\
&&\quad={\Pr}[ Y\in B_1-\delta|\|\delta_0-\delta\| ]\nonumber
\\\label{mumeas}
&&\quad=\mathrm{E}\{ {\Pr}[ \vec Y\in\|Y\|^{-1}(B_1-\delta)|\|Y\|,\|\delta_0-\delta\| ]|\nonumber
\\[-8pt]\\[-8pt]
&&\hspace*{32pt}\|\delta_0-\delta\|\}\nonumber
\\
&&\quad=\mathrm{E}\bigl\{ \mu_p\bigl(\|Y\|^{-1}(B_1-\delta)\bigr)|\|\delta_0-\delta\| \bigr\}.\nonumber
\end{eqnarray}
Because $B_1-\delta$ is a ball with $0\in\partial(B_1-\delta)$, the set
$(\|Y\|^{-1}(B_1-\delta))\cap\mathcal{S}_p$ is a
spherical cap on
$\mathcal{S}_p$ which, after some geometry, can be expressed as
%
\begin{equation}\label{cap}
\biggl\{(z_1,\dots,z_p)\Bigl|\frac{z_1}{(z_1^2+\cdots+z_p^2)^{1/2}}
\ge
\frac{\|Y\|}{2\|\delta_0-\delta\|}\biggr\}\hspace*{-27pt}
\end{equation}
when $\|Y\|\le2\|\delta_0-\delta\|$, and is empty otherwise.
Furthermore, $\mu_p$ can be
represented as the distribution of $\vec Z$, where
$Z\equiv(Z_1,\dots,Z_p)\sim N_p(0,1_p)$. Therefore,
%
\begin{eqnarray}\label{muprob}
&&\mu_p\bigl(\|Y\|^{-1}(B_1-\delta)\bigr)\nonumber
\\
&&\quad=\frac{1}{2}{\Pr}\biggl[\frac{Z_1^2}{Z_1^2+\cdots+Z_p^2}\ge\frac{\|Y\|^2}{2\|\delta_0-\delta\|^2}\Bigl|
\\
&&\hspace*{104pt}\quad
\|Y\|, \|\delta_0-\delta\|\biggr]\nonumber
\end{eqnarray}
when $\|Y\|\le2\|\delta_0-\delta\|$, and $= 0$ otherwise. Thus, by~\eqref{mumeas},
%
\begin{eqnarray}\label{idontknow}
&&{\Pr}_\delta[ X\in B_1|\|\delta_0-\delta\| ]\nonumber
\\
&&\quad\le\frac{1}{2}{\Pr}\biggl[\frac{Z_1^2}{Z_1^2+\cdots+Z_p^2}
\ge\frac{\|Y\|^2}{2\|\delta_0-\delta\|^2}\Bigl|
\\
&&\hspace*{127pt}\quad\|\delta_0-\delta\| \biggr],\nonumber
\end{eqnarray}
hence,
\begin{eqnarray}\label{idontknow2}
&&{\Pr}_\delta[ X\in B_1] \nonumber
\\[-8pt]\\[-8pt]
&&\quad\le\frac{1}{2}{\Pr}\biggl[\frac{Z_1^2}{Z_1^2+\cdots+Z_p^2}
\ge\frac{\|Y\|^2}{2\|\delta_0-\delta\|^2}\biggr].\nonumber
\end{eqnarray}
But $Z_1^2+\cdots+Z_p^2=O(p)$ in probability by the Law of Large Numbers,
so by \textup{(iii)}, the right-hand side of~\eqref{idontknow2} approaches 0 as
$p\to\infty$, which yields \eqref{newlimit5}.
\end{pf}
%

\renewcommand{\thefigure}{\arabic{figure}}
\setcounter{figure}{3}
\begin{figure*}
\begin{tabular}{@{}c@{\quad}c@{}}

\includegraphics{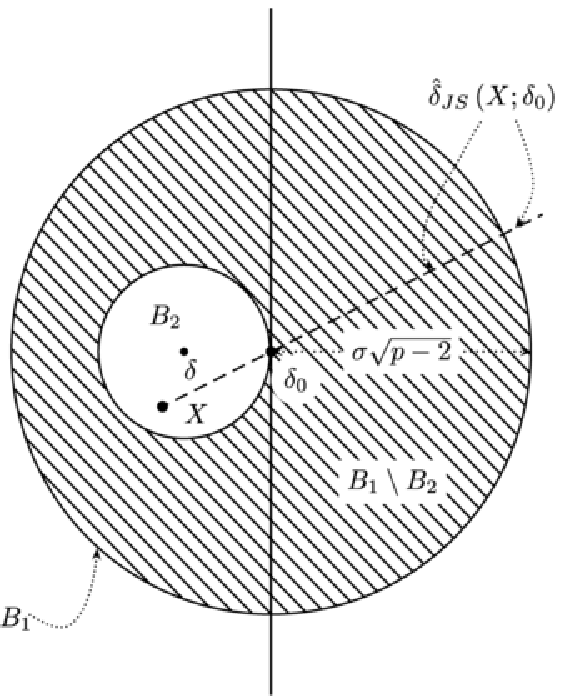}
&\includegraphics{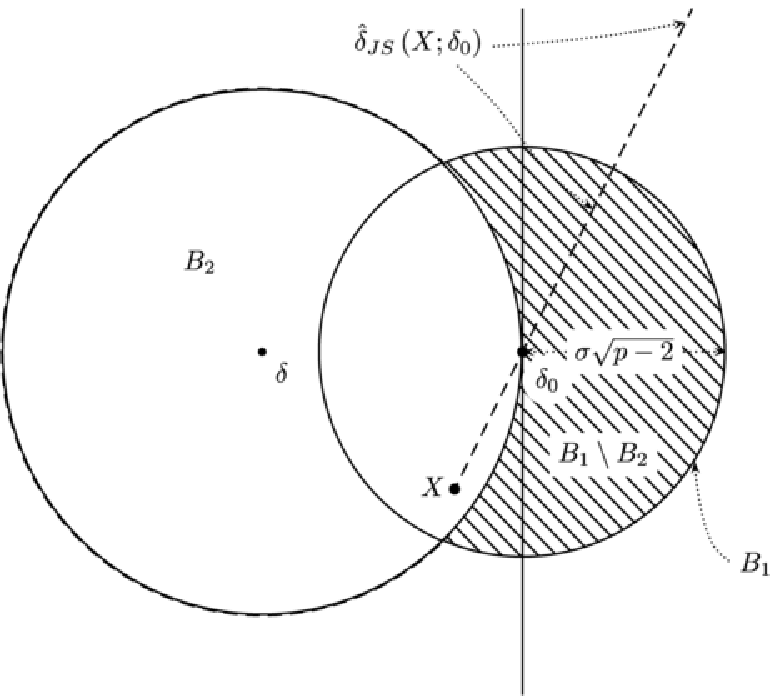}\\
(a) $\|\delta_0-\delta\|<\frac{\sigma}{2}\sqrt{p-2}$&
(b) $\|\delta_0-\delta\|>\frac{\sigma}{2}\sqrt{p-2}$
\end{tabular}
\caption{Illustrating the implication \protect\eqref{implies}.}\label{fig:4}
\end{figure*}

\section{\texorpdfstring{Verification of (\protect\ref{S2})}{Verification of (5)}}
\label{appendix2}

\setcounter{equation}{44}
If we set $h(\gamma)=\|\hat \delta_\gamma(X;\delta_0)-\delta\|^2$ and
$\bar\delta=\frac{1}{2}(\delta_0+\delta)$, then
\begin{eqnarray}\label{hderiv}
h'(1)&=&2(X-\delta_0)^t(X-\delta)\nonumber
\\[-8pt]\\[-8pt]
&=&2[\|X-\bar\delta\|^2-\|\delta-\bar\delta\|^2].\nonumber
\end{eqnarray}
Since the right-hand side of \eqref{S2} is the set $\{X\mid h'(1)>0\}$, \eqref{S2} follows.
%

\section{\texorpdfstring{Verification of (\protect\ref{S12})}{Verification of (12)}}
\label{appendix3}

\setcounter{equation}{45}
First note that $\hat \delta_{\mathrm{JS}}(X;\delta_0)\ne\hat \delta_{\mathrm{JS}}^+(X;\delta_0)$ iff
$\|X-\delta_0\|<\sigma\sqrt{p-2}$ (the ball $B_1$ of radius $\sigma\sqrt{p-2}$ centered at
$\delta_0$; see Figures~\ref{fig:4}a,~\ref{fig:4}b), in which case
$\hat \delta_{\mathrm{JS}}^+(X;\break\delta_0)=\delta_0$.
Define
%
\begin{eqnarray}\label{setC}
C&:=&\{X|\|\hat \delta_{\mathrm{JS}}^+(X;\delta_0)-\delta\|<\|X-\delta\|\},
\\\label{setD}
D&:=&\{X|\|\hat \delta_{\mathrm{JS}}(X;\delta_0)-\delta\|<\|X-\delta\|\},
\end{eqnarray}
so \eqref{S12} is equivalent to
%
\begin{equation}\label{CgreaterD}
{\Pr}[C\mid\delta_0]>{\Pr}[D\mid\delta_0].
\end{equation}
Since $C\setminus B_1=D\setminus B_1$, this is equivalent to
%
\begin{equation}\label{CgreaterDcapB1}
{\Pr}[C\cap B_1\mid\delta_0]>{\Pr}[D\cap B_1\mid\delta_0].
\end{equation}
But (see Figures~\ref{fig:4}a,~\ref{fig:4}b)
%
\begin{equation}\label{implies}
X\in B_2\quad\Rightarrow\quad\|\hat \delta_{\mathrm{JS}}(X;\delta_0)-\delta\|>\|X-\delta\|,\hspace*{-20pt}
\end{equation}
where $B_2$ is the ball of radius $\|\delta_0-\delta\|$ centered at~$\delta$, so $D\cap B_2=\varnothing$. Thus,
%
\begin{equation}\label{inclusion}
C\cap B_1=(B_1\setminus B_2)\supsetneq D\cap B_1,
\end{equation}
hence, \eqref{CgreaterDcapB1} holds. (Note that no distributional
assumption on $X$ is
needed.)

\section{\texorpdfstring{Verification of Figures~\protect\ref{S3333}a and~\protect\ref{S3333}b; Verification of (\protect\ref{S19})}
{Verification of Figures 3a and 3b; Verification of (19)}}
\label{appendix4}

\setcounter{equation}{51}
First we verify that Figures~\ref{S3333}a and~\ref{S3333}b accurately depict the region
%
\begin{equation}\label{Fig3region}
\{\delta_0|\|\hat \delta^+_{\mathrm{JS}}(X;\delta_0)-\delta\|<\|
X-\delta\|\}.\vadjust{\goodbreak}
\end{equation}
Let $\gamma=(1-\frac{\sigma^2(p-2)}{\|X-\delta_0\|^2} )^+$, so
$0\le\gamma<1$ and $\hat \delta_{\mathrm{JS}}^+(X;\delta_0)=\gamma(X-\delta_0)+\delta_0$. Each of the
following inequalities is equivalent to that in
\eqref{Fig3region}:
\begin{eqnarray*}
&&\|(1-\gamma)(\delta_0-\delta)+\gamma(X-\delta)\|^2<\|X-\delta\|^2,
\\
&&(1-\gamma)^2\|\delta_0-\delta\|^2+2\gamma(1-\gamma)(\delta_0-\delta)^t(X-\delta)
\\
&&\quad<(1-\gamma^2)\|X-\delta\|^2,
\\
&&(1-\gamma)\|\delta_0-\delta\|^2+2\gamma(\delta_0-\delta)^t(X-\delta)
\\
&&\quad<(1+\gamma)\|X-\delta\|^2,
\\
&&\|\delta_0-\delta\|^2<\gamma[\|X-\delta_0\|^2]+\|X-\delta\|^2.
\end{eqnarray*}
If $\|\delta_0-X\|^2<\sigma^2(p-2)$, that is, $\delta_0$ lies inside
the ball of radius
$\sigma\sqrt{p-2}$ (see Figures~\ref{S3333}a,~\ref{S3333}b), then $\gamma=0$ and the
last inequality becomes
$\|\delta_0-\delta\|^2<\|X-\delta\|^2$, which holds iff $\delta_0$
lies inside the ball $B$
of radius $\|X-\delta\|$ centered at $\delta$. If $\|\delta_0-X\|^2<\sigma^2(p-2)$, that is,
$\delta_0$ lies outside this ball, then $\gamma=(1-\frac{\sigma^2(p-2)}{\|X-\delta_0\|^2} )$
and the last inequality instead is equivalent to each of the following:
\begin{eqnarray*}
\|\delta_0-\delta\|^2&<&\|X-\delta_0\|^2+\|X-\delta\|^2-\sigma^2(p-2),
\\
\sigma^2(p-2)&<&2\|X-\delta\|^2+2(X-\delta)^t(\delta-\delta_0),
\\
\sigma^2(p-2)&<&2(X-\delta)^t(X-\delta_0),
\\
\frac{\sigma^2(p-2)}{2\|X-\delta\|}&<&\overrightarrow{(X-\delta)}^t(X-\delta_0),
\end{eqnarray*}
which holds exactly in the open halfspace $K$ shown in Figures~\ref{S3333}a and~\ref{S3333}b.
Thus, the region \eqref{Fig3region} is the union $B\cup K$ of the cross-hatched regions
in these figures.\vadjust{\goodbreak}

Finally, we verify \eqref{S19}, which now can be written
equivalently as
%
\begin{equation}\label{pcloser7}
\lim_{p\to\infty}{\Pr}_\delta[ \delta_0\in B\cup K]=0,
\end{equation}
by the following proposition, in which $\delta$, $\delta_0$, $X$,
$V$,~$\sigma$, and
$\tau$ now depend on $p$.

\begin{proposition}\label{prop3}
Assume the following:
\begin{longlist}[(iii$^{\prime}$)]
\item[(i$^{\prime}$)] $V\equiv\delta_0-X$ is independent of $X$;
\item[(ii$^{\prime}$)] $V$ is spherically symmetric;\vspace*{1pt}
\item[(iii$^{\prime}$)]
$\frac{\|X-\delta\|}{\|V\|}\equiv
\frac{\|X-\delta\|}{\|\delta_0-X\|}=o(p^{1/2})$ in probability as $p\to\infty;$
\item[(iv$^{\prime}$)] $\sigma^{-2}\|X-\delta\|\cdot\|V\|\equiv
\sigma^{-2}\|X-\delta\|\cdot\|\delta_0-X\|=o(p^{3/2}) \mbox{ in probability as } p\to\infty$.
Then [cf. \eqref{S19}]
%
\begin{equation}\label{pcloser10}
\hspace*{20pt}\lim_{p\to\infty}{\Pr}_\delta[ \|\hat \delta^+_{\mathrm{JS}}(X;\delta_0)-\delta\|>\|X-\delta\| ]=1.
\end{equation}
The boundedness assumption \textup{(iii$^{\prime}$)} [resp., \textup{(iv$^{\prime}$)}] is
satisfied, for example, if
$X\sim N_p(\delta,\sigma^21_p)$ and $\delta_0\sim N_p(X,\tau^21_p)$ with
$\tau/\sigma=o(p^{1/2})$ [resp., $\sigma/\tau= o(p^{1/2})$], so both
are satisfied if
$\tau/\sigma\sim p^\varepsilon$ with $0\le|\varepsilon|<1/2$.
\end{longlist}
\end{proposition}

\begin{pf}
By the argument that yielded \eqref{newlimit5}
in Appendix~\ref{appendix1} [with \textup{(i)}--\textup{(iii)} and $X$, $Y$, $B_1-\delta$, and
$\|\delta_0-\delta\|$ replaced
by \textup{(i$^{\prime}$)}--\textup{(iii$^{\prime}$)} and $\delta_0$, $V$, $B-X$, and $\|X-\delta\|$],
we obtain
%
\begin{equation}\label{pcloser8}
\lim_{p\to\infty}{\Pr}_\delta[ \delta_0\in B ]=\lim_{p\to \infty}{\Pr}_\delta[ V\in B-X ]=0.\hspace*{-27pt}
\end{equation}
Next, again by the argument in Appendix~\ref{appendix1} but with $B_1-\delta$
replaced by $K-X$,
%
\begin{eqnarray}
\hspace*{24pt}&&{\Pr}_\delta[ \delta_0\in K|\|X-\delta\| ]
\\
&&\quad=\mathrm{E}\bigl\{ \mu_p\bigl(\|V\|^{-1}(K-X)\bigr)|\|X-\delta\| \bigr\}
\\
&&\quad=\frac{1}{2}{\Pr}\biggl[\frac{Z_1^2}{Z_1^2+\cdots+Z_p^2}
\ge\frac{\sigma^4(p-2)^2}{4\|X-\delta\|^2\|V\|^2}\Big|\nonumber
\\[-8pt]\\[-8pt]
&&\hspace*{153pt}\quad
\|X-\delta\| \biggr].\nonumber
\end{eqnarray}
Thus, by \textup{(iv$^{\prime}$)},
%
\begin{eqnarray}\label{pcloser9}
\hspace*{23pt}&&\lim_{p\to\infty}{\Pr}_\delta[ \delta_0\in K]\nonumber
\\
&&\quad=\frac{1}{2}\lim_{p\to\infty}{\Pr}
\biggl[\frac{Z_1^2}{Z_1^2+\cdots +Z_p^2}
\\
&&\hspace*{60pt}\quad\ge\frac{\sigma^4(p-2)^2}{4\|X-\delta\|^2\|V\|^2}\biggr]=0,\nonumber
\end{eqnarray}
so \eqref{pcloser7} and \eqref{pcloser10} are confirmed.
\end{pf}
\end{appendix}

\section*{Acknowledgments} We gratefully acknowledge the
contributions of T.~W. Anderson, Steen Andersson, Morris Eaton, Charlie Geyer, Erich Lehmann,
Ingram Olkin, and Charles Stein to our
understanding of the role of invariance in statistical analysis. We
warmly thank Mathias Drton, Brad Efron, Carl Morris, and Jon Wellner for helpful comments
and suggestions. This research was supported in part by NSA Grant
MSPF-05G-014 and Grant R-155-000-081-112 from the National University of Singapore.

\end{document}